\documentclass{aastex}
\shortauthors{Alpar}
\shorttitle{Pulsar Braking Indices}
\begin{document}
\title{Pulsar Braking Indices, Glitches and Energy Dissipation In 
Neutron Stars}
\vspace{3cm}
\author{M.Ali Alpar$^{1}$, Altan Baykal$^{2}$ }
\affil {
$^{1}$ Sabanc{\i} University, Istanbul,
Turkey\\
$^{2}$ Middle East Technical University, Physics Department, 
Ankara, Turkey}


\begin{abstract}

Almost all pulsars with anomalous positive $\ddot \Omega $ measurements 
(corresponding to anomalous braking indices in the range
5$<n<$100), including all the pulsars with 
observed large glitches ($\Delta\Omega/\Omega$ $>$ 10$^{-7}$) 
as well as post glitch or interglitch 
$\ddot \Omega$ measurements obey the scaling between 
$\ddot \Omega$ and glitch parameters originally noted in the Vela pulsar. 
Negative second derivative values
can be understood in terms of glitches that were missed or remained
unresolved. We discuss the glitch rates and a priori probabilities of 
positive and negative braking indices according to the model developed 
for the Vela pulsar. This behavior supports the universal occurrence of a 
nonlinear dynamical coupling between the neutron star crust and an interior 
superfluid component. The implied lower limit to dynamical energy dissipation 
in a neutron star with spindown rate $\dot \Omega$ is 
$\dot E_{diss}> 1.7 \times 10 ^{-6} \dot E_{rot}$. 
Thermal luminosities and surface temperatures due to 
dynamical energy dissipation are estimated for old neutron stars which are 
spinning down as rotating magnetic dipoles beyond the pulsar death line.

\end{abstract}
 
\newpage
\section{Introduction}

Anomalous second derivatives of the rotation rates 
of radio pulsars may have interesting implications.
 Very large positive or negative second 
derivatives are likely to be artefacts of timing noise. We show here that 
second derivatives corresponding to braking indices n in the 
interval 5 $<$ $\mid $n$\mid$ $<$ 100 generally 
fit well with secular interglitch behaviour 
according to a model previously applied to the Vela pulsar. 
Pulsars with large glitches ($\Delta\Omega/\Omega$ $\ge$ 10$ ^{-7}$) and 
measured anomalous second derivatives of the rotation rate, mostly positive 
(Shemar $\&$ Lyne 1996, Lyne, Shemar $\&$ Graham-Smith 2000, Wang et al. 2000), 
as well as pulsars with positive or negative anomalous second derivatives 
but no observed glitches (Johnston $\&$ Galloway 1999) scale with the model.
We infer that isolated neutron stars older than Vela have dynamical 
behaviour similar to the Vela pulsar. This implies relatively large energy 
dissipation rates that can 
supply a luminosity to older isolated neutron stars. 

The spindown law of a pulsar is usually given in the form 
 $\dot \Omega = -k \Omega ^{n}$ where n, the braking index, is 3 if the 
pulsar spindown is determined purely by electromagnetic radiation
torques generated by the rotating magnetic dipole moment of the neutron 
star. The braking index has been conventionally 
measured through the relation
\begin{equation}
n=\frac{\Omega \ddot \Omega}{\dot \Omega^{2}}
\end{equation}
by measuring $\ddot \Omega$, the second derivative of the pulsar rotation 
frequency. An alternative method, suggested recently 
by Johnston $\&$ Galloway (1999) is based on integrating, rather than 
differentiating, the spindown law, to obtain 
\begin{equation}
n=1+\frac{ \Omega_{1} \dot \Omega_{2}-\Omega_{2} \dot \Omega_{1}}
{\dot \Omega_{1} \dot \Omega_{2} (t_{2}-t_{1})} 
\end{equation}
where $\Omega _{i}$ and $\dot \Omega _{i}$ are values measured at $t_{i}$.   

Among the radio pulsars known, only young pulsars have braking
indices measured with accuracy. These reported braking indices are all 
less than 3: 
For the Crab pulsar
n= 2.509 $\pm$ 0.001 (Lyne, Pritchard $\&$ Smith 1988, Lyne, Pritchard
$\&$ Smith 1993); for PSR B 1509-58, n= 2.837 $\pm$ 0.001
(Kaspi et al. 1994); for PSR B 0540-69, n= 2.04 $\pm$ 0.02 (Manchester
$\&$ Peterson 1989, Nagase et al. 1990, Gouiffes, Finley $\&$ \"Ogelman
1992); for pulsar J 1119-6127, n= 2.91 $\pm$ 0.05 (Camilo et al. 2000); for 
pulsar J 1846-0258, n=2.65 $\pm$ 0.01 (Livingstone et al. 2006). 
For the Vela pulsar a long term (secular) 
braking index of 1.4 $\pm$ 0.2 was reported (Lyne et al. 1996).
This value was extracted with certain assumptions for connecting fiducial 
epochs across a timing history dominated by glitches and interglitch response.

For old pulsars with $\nu \sim 1$ Hz 
and $\dot \nu \sim 10^{-15}$ Hz s$^{-1}$, the expected
$\ddot \nu$ for n=3 is $\sim 10^{-30}$ Hz $s^{-2}$. 
This is difficult to measure because
the cumulative effect of the second derivative would contribute one extra 
cycle count ($ (\ddot \nu t^{3})/6 \sim 1$) 
only after several centuries.
For 19 "old" radio pulsars, observations yielded 
anomalous braking indices extending from $\sim \pm 4$ all the way 
to $\pm 10^{5}$ (Gullahorn $\&$ Rankin 1982). Later
measurements of braking indices of these pulsars have
shown that these anomalous values are artefacts 
produced by timing noise (Cordes 1980, Cordes $\&$ Helfand 1980, 
Cordes $\&$ Downs 1985).
Some of the old pulsars' 
(PSRs 0823+26, 1706-16, 1749-28, 2021+51)
 time of arrival (ToA) data
extending over more than three decades were investigated for the stability
of the pulse frequency second derivatives $\ddot \nu $
(Baykal et al. 1999).
These pulsars have shown anomalous values of braking indices of the 
order of $\sim$ $\pm$ 10$^{5}$. In the framework of low resolution
noise power spectra estimated from the residuals of pulse
frequency and ToA data, it is found (Baykal et al. 1999) that the
$\ddot \nu $ terms of these sources arise from the
red torque noise in pulse frequency derivatives.

For pulsars with moderate ages, $\sim 10^{5}$ yr, 
anomalous braking indices have values of order  $\pm 10^{2}$.
These are not noise artefacts. Rather, such braking indices can be understood 
as part of the neutron star's secular dynamics. 
The interglitch recovery of pulsars extending 
through observation time spans may yield positive anomalous braking indices, 
while negative anomalous braking indices can be explained 
by the occurrence of an unobserved glitch causing 
a negative step $\Delta \dot \Omega$ in the spindown rate 
(as typically observed with resolved glitches),
between the different measurements of $\dot \Omega$ (Johnston \& Galloway 1999).
In this work we show that all pulsars with
anomalous $\ddot \Omega $ measurements, including all the pulsars with
observed glitches as well as post glitch or interglitch
$\ddot \Omega$ values (Shemar \& Lyne 1996, Wang et al. 1999) 
obey the same scaling between $\ddot \Omega$ and glitch
 parameters (Alpar 1998) as in the models
 developed for the Vela pulsar glitches (Alpar et al. 1993).

The prototypical Vela pulsar glitches occur at intervals of about 2 years. 
Models developed for the Vela pulsar glitches indicate that interglitch 
intervals scale with $|\dot \Omega|^{-1}$. This is borne out by 
the statistics of large $\Delta\Omega/\Omega > 10^{-7}$ glitches 
(Alpar $\&$ Baykal 1994). Scaling with the spindown rates, the glitch
intervals of pulsars at the ages of $10^{5}-10^{6}$ yr are
of the order of $\sim 10^{2}$ yrs.

In Section 2, we review the observations of anomalous braking indices,  
their errors and methods of deciding if the nominal second derivatives are 
artefacts of the noise process.
In Section 3, we review the interglitch timing behaviour of the Vela pulsar
 and the simple explanation for this standard behaviour in terms of the
 model of nonlinear vortex creep dynamics in the neutron star superfluid. 
In Section 4, we show that pulsars with reliable anomalous 
$\ddot \Omega $ measurements
 can be consistently explained within the same model, 
with one model parameter whose values are similar, 
to order of magnitude, to those 
obtained in detailed fits to the Vela pulsar timing data.
In Section 5, we extend this analysis to 
 pulsars with glitches 
of size $\Delta \Omega / \Omega > $10$^{-7}$, comparable to the Vela
pulsar glitches and with reliable anomalous 
$\ddot \Omega $ measurements. 
This seemingly universal dynamics is characterized by a lag in rotation 
rate between the observed crust and some interior component 
of the neutron star, the crust superfluid in current models. 
The identification of the universal
 dynamical behaviour leads us to derive a lower 
limit on the lag, and a corresponding lower limit on the rate of 
dynamical energy dissipation.
In Section 5, we explore the implications of the lower bound on the 
energy dissipation rate. Estimates of minimum thermal luminosities 
and surface blackbody temperatures for isolated neutron stars 
of various ages are presented under the dipole spindown law. 

\section{Observations of Anomalous Braking Indices}

Pulse arrival time measurements display
 irregularities in the rotation rate
known as "timing noise".
The timing noise
could be due to a noisy component of the secular torque
involving fluctuations in the magnetosphere
of the neutron star (Cheng 1987 a,b; 1989).
Alternatively, timing noise could arise
from internal torques coupling different components of the
neutron star, for example the decoupling and recoupling of the crust superfluid
(Alpar, Nandkumar $\&$ Pines 1986,
Jones 1990).
Timing noise for pulsars has been studied for the last three decades
(Boynton et al. 1972, Groth 1975,
Cordes 1980, Cordes $\&$ Helfand 1980,
 Cordes $\&$ Downs 1985, D'Alessandro et al. 1995, 1997,
Deshpande et al. 1996).
Boynton et al. (1972) proposed that the timing noise in the
times of arrival (ToA) of pulses
might arise from "random walk" processes which are
r$^{th}$ order (r=1,2,3)
time integrals
of a 'white noise' time series
 (that is, a time series of unresolved delta functions).
The random walks in phase $\phi $, pulse frequency $\nu$ and
pulse frequency derivative $\dot \nu$ are called
"phase noise",
"frequency noise" and "slowing down noise" respectively
(Cordes 1980).

The cross-talk between the
timing noise and secular slowing down is very important.
 Many of the old pulsars with spin-down age
$\tau = P/2\dot P$ greater than about
10$^{6}$ years have shown
anomalous trends in their secular frequency second derivative ($\ddot \nu$)
(Cordes $\&$ Downs 1985).
 These
trends make it impossible to recover the braking law
$\dot \nu \sim  \nu ^{n} $ of the pulsar
(for pure magnetic dipole radiation n=3).
 Nominal values of $\ddot \nu$ from timing fits gave anomalous
 braking indices ranging from $-10^{5}$ to $10^{5}$ in various pulsars.
 Recent observations
of some young/middle aged pulsars with glitches also
showed anomalous
positive braking indices of the order $\sim $20 - 200 (Shemar $\&$ Lyne 1996, 
Lyne, Shemar $\&$ Graham-Smith 2000, Wang et al. 2000).
 Interglitch recovery between successive glitches can effect
the pulsar's dynamical parameters such as $\dot \nu$ $\&$ $\ddot \nu $.
For the glitching pulsars, the high values of the second derivative
of the rotation rate, $\ddot \nu $,
and associated braking indices of order 20-200
are characteristic of interglitch
recovery
 (Alpar 1998),
which extends from one glitch to the next one,
as studied in detail between the glitches of the Vela
pulsar (Alpar et al. 1993).
For all middle aged pulsars the expected intervals between glitch events are of
the
order of a few hundred years
 (Alpar $\&$ Baykal 1994).
Thus a pulsar is most likely to be observed during the
interglitch recovery phase.
A sample of pulsars without observed glitches (Johnston $\&$ Galloway 1999) 
displays mostly positive, along with some negative braking indices. 

Baykal et al. (1999) have investigated the time series of
pulsars on the longest available time scales by combining 
observations of 24 pulsars (Downs $\&$ Reichley 1983) with
later observations (Siegman, Manchester $\&$ Durdin 1993,
Arzoumanian, Nice $\&$ Taylor 1994)
 containing available timing data for time spans of the order of 30 years
for several pulsars.
Some of these pulsars were eliminated
as candidates for secular timing behaviour, since their
frequency time series are not consistent with secular quadratic trends
(constant $\ddot \nu$). Equivalently, polynomial fits to the ToA of
these pulsars 
require higher order polynomials rather than a cubic polynomial.
For these pulsars the time series is dominated by complicated
noise processes rather than interglitch recovery.
For four pulsars, 
PSR 0823+26, PSR 1706-16, PSR 1749-28, and
 PSR 2021+51, the time series called for a more careful analysis to 
determine if there is a secular second derivative. While  
there are significant quadratic trends in frequency histories
(cubics in ToA), these trends arise from the cumulative effect of 
noise.
Baykal et al. (1999) estimated the noise strengths for 
these four pulsars from the residuals
of ToA data.
In order to see whether the noise strengths are
stable or not and to see whether the
quadratic trends in pulse frequency
and
cubic trends in ToA absorb the noise, they estimated alternative sets of
noise strengths by removing quadratic polynomials
from the pulse frequency data for the longest time span of data
and cubic polynomials
from the ToA data for the shorter intervals. They  found that
for each source these two
power spectra are consistent with each other
in terms of average noise strength
$S_{r}$ and slope of the power spectra.
This suggested that their original noise estimates
 were robust,
(consistent with each other in terms of the noise strength parameter,
$S_{r}$) and
were not dominated by either of the two particular polynomial trends.
If there were a secular polynomial trend in the data, one would 
expect that particular polynomial trend to produce a significantly better fit,
 i.e. a significantly lower, and different, power spectrum 
of the residuals, compared to the
other polynomial models. All pulsars investigated by Baykal et al. 
(1999) are old pulsars, with characteristic ages $P/2\dot P > 10^{7}$ yrs. 

In the technique developed by Johnston $\&$ Galloway (1999) the
braking index is obtained from
$\nu$ and $\dot \nu$ values. Errors of braking indices
depend on the errors of $\nu$ and $\dot \nu$. Johnston 
and Galloway applied their methods
to 20 pulsars. They found that the braking indices of old pulsars 
are insignificant because of large error bars. 
However pulsars with middle ages have yielded significant braking 
indices. Due to the sparseness of timing data power spectrum 
techniques cannot be applied to these pulsars. 
All "middle aged" ($10^{5}$ yrs$~<\tau<~10^{7}$ yrs) and young 
pulsars have large spindown rates compared to the spindown rates 
of old pulsars. Observations of anomalous braking indices 
suggested that the old pulsars' braking indices are artefacts of
timing noise. For the young and middle aged pulsars
timing noise does not have a strong effect on $\ddot \nu$ values. 
In this work we take the young and middle aged pulsars' braking
indices to be real and older pulsars' braking indices to be artefacts of 
timing noise. This is in agreement with the result of 
Johnston $\&$ Galloway (1999),
on the basis of data from 20 pulsars,
 and with the results of
 Baykal et al. (1999) for four old pulsars.

\section{The Model for Glitches and Interglitch Dynamics }

Extensive timing observations on the Vela pulsar now cover 
a period of about 35 years and encompass 14 glitches with 
postglitch relaxation and interglitch timing behaviour. A detailed 
empirical model interprets the glitches and post glitch-interglitch 
response in terms of angular momentum exchange between a "pinned 
crust superfluid" and the observed crust of the pulsar 
(Alpar et al. 1984a,b, 1989, 1993). The time t$_{g}$ 
between glitches scales as $ \mid \dot \Omega \mid ^{-1}$ in this model.
The hypothesis that all pulsars experience glitches similar to the 
Vela pulsar glitches, at rates proportional to the $ \mid \dot \Omega \mid$ 
of the individual pulsars, is borne out by the statistics of Vela 
type ($\Delta \Omega / \Omega > $10$^{-7}$) glitches from the entire 
pulsar sample (Alpar $\&$ Baykal 1994). The observations 
of glitches and interglitch measurements 
of $\ddot \Omega$ (Shemar $\&$ Lyne 1996, Johnston $\&$ Galloway 1999, 
Lyne, Shemar $\&$ Graham-Smith 2000, Wang et al. 2000) provide us 
with many pulsars actually observed in behaviour like the Vela pulsar 
prototype. Our first task is to demonstrate this similarity in 
dynamical behaviour. We start with a summary of the 
model developed for the Vela pulsar. The basic features will be brought 
forth in a description involving the observed neutron star crust and one
 interior component, and independently of the microsopic details
 of the coupling between the two components. 

In the absence of evidence that the pulsar electromagnetic torque 
changes at a glitch, and with the established impossibility
 of explaining the large ($\Delta \Omega /\Omega >10^{-7}$) and frequent 
(intervals $\sim$ 2 yrs) Vela pulsar glitches with starquakes, 
the glitch is modelled as a sudden angular momentum exchange
 between the neutron star crust and an interior component, 
\begin{equation}
I_{c}\Delta \Omega _{c} = I_{s}\delta \Omega = (I_{A}/2 + I_{B}) \delta \Omega. 
\end{equation}   
Here $\Delta \Omega _{c}$ is the observed increase of the crust's rotation
 rate at the glitch. $I_{c}$ is the effective moment of inertia of the crust, 
including all components of the star dynamically coupled to the crust 
on timescales shorter than the resolution of the glitch event.
 The observations imply that $I_{c}$ includes practically
 the entire moment of inertia of the star, and the theory of the
 dynamical coupling mechanisms of the neutron star core
 (Alpar, Langer $\&$ Sauls 1984) provides an understanding
 of this by furnishing crust-core coupling times shorter
 than the resolution of glitch observations. 
 The coupling mechanism relies on the simultaneous presence of 
superfluid neutrons and superconducting protons in the
core of the neutron star. Recent arguments that the protons in the core 
of the neutron star are either normal or in the Type
I superconductor phase would kill this coupling mechanism for all or 
some regions of the core neutron superfluid, which
carries almost the entire moment of inertia. This would then require 
another mechanism of short time scale coupling of the
core superfluid neutrons to the effective crust to explain the empirical 
fact that almost the entire moment of inertia of the
neutron star seems to couple to the observed crust rotation on 
timescales less than a minute. However, the argument for the
absence of Type II proton superconductivity is not strictly valid 
because it rests on the premises that (i) the observed long
term modulation in timing and pulse shapes of the pulsar PSR B1828-11 is 
due to precession of the neutron star and not due to
some surface or magnetospheric excursion of the magnetic field pattern, 
and (ii) that such precession of the observed period
and amplitude cannot take place in the presence of pinning. Of these 
premises (i) is not necessarily the case, and (ii) is
not valid because at finite temperature pinning does not give an 
absolute constraint on precession (Alpar 2005).
In the following we replace $I_{c}$ with $I$, the total moment of inertia 
of the star. 

In 
 current models the sudden transfer of angular momentum is 
associated with a superfluid in the inner crust of the neutron star, 
where the rotational dynamics of the superfluid is constrained by the 
existence of pinning forces exerted by the crust lattice on the superfluid's vortex lines. 
$\delta \Omega$ describes the decrease in the rotation rate of the pinned
 superfluid at the  glitch. $I_{A}$ and $I_{B}$ are parts of the superfluid's
 effective moment of inertia $I_{s}$ associated with different dynamical behaviour.
 The vortex lines are the discrete carriers of the superfluid's angular momentum.
 Vortex lines under pinning forces respond to the driving external pulsar torque,
 as this torque makes the normal crust lattice spin down.
 
There are two modes of this response. Some vortices will remain pinned until
 critical conditions matching the maximum available pinning force are reached.
 Then they will unpin catastrophically and move rapidly in the radially
 outward direction, thereby transferring angular momentum to the crust
 only in glitches.
 The element of the superfluid through which unpinned 
vortices move rapidly in a glitch, and there is no vortex flow 
otherwise, has moment of inertia $I_{B}$ and contributes angular
momentum $I_{B} \delta \Omega $ to the glitch in rotation frequency, as 
indicated in Eq. (3).
It does not spin down continuously between glitches, rather it spins down only
 by discrete steps of the angular momentum transfer at glitches,
 analogous to a capacitor which does not transmit electric current except
 in discharges. Since it does not contribute to spindown between glitches,
 it does not contribute to the glitch induced sudden change in spindown rate. 

 In other parts of the superfluid, vortices are not pinned all the time,
but unpin and repin, at thermally supported rates.
$I_{A}$ is the moment of inertia of
 those parts of the superfluid that allow a continuous
vortex flow, in analogy with the current in a resistive circuit element.
In the presence of finite energy barriers, there will always be
a continuous current of vortices, in addition to the discrete discharges that
we call glitches. This continuous current of vortices moving radially
outward through the inner crust "vortex creep" makes the superfluid spin down
continously in
response to the driving spindown torque on the pulsar.
At finite temperature, the motion of these vortices
against the pinning energy barriers is made possible by thermal activation.
A different possibility, operating even at T = 0, is quantum tunneling.
 It can be shown easily that if vortices unpinned in a glitch are
unpinned at a uniform density throughout the creep regions
of moment of inertia  $I_{A}$, then the angular momentum transfer from these
regions to the normal crust is $ I_{A} \delta \Omega /2 $,
as in the right hand side of Eq. (3) (Alpar et al. 1984a, 1993).

The continuous spindown between glitches is governed by:
\begin{equation}
I_{c}\dot\Omega _{c} = N_{ext} + N_{int} = N_{ext} - I_{A}\dot\Omega _{s}, 
\end{equation}
where $N_{ext}$ is the external torque on the neutron star, 
and $N_{int}$ is the internal torque coupling the 
superfluid to the "effective crust" with moment of inertia 
$I_{c} \cong I$.
 
 In a cylindrically symmetric situation the spindown rate of the superfluid
 is proportional to the mean vortex velocity in the radial direction, which in turn is determined 
by the lag $\omega = \Omega - \Omega_c$ between the superfluid
and crust rotation rates:
\begin{equation}
\dot \Omega_{s}=-\frac{2\Omega_{o}}{r}V_{r} (\omega).
\end{equation}
As the glitch imposes a sudden change in $\omega$, it will offset 
the superfluid spindown and therefore the observed spindown rate of the crust,
 according to Eq. (4). The glitch is followed by transient relaxation processes
 in which the crust rotation frequency and spindown rate relax promptly as an
 exponential function of time (Alpar et al. 1984a,b). It is the long term interglitch
 relaxation of the spindown rate, after the transients are over, that determines
 the interglitch behavior of the observed crust spindown rate.
 Labelling the moment of inertia associated with long term offset
 in spindown rate with $I_{A}$, from Eq.(4) we have 
\begin{equation}
\frac{\Delta \dot \Omega}{\dot \Omega}=\frac{I_{A}}{I}.
\end{equation}
We refer the reader to earlier papers (Alpar et al. 1984 a,b, 1989) for details. The
contribution of the regions $I_{A}$ to the glitch in the rotation frequency is
$I_{A} \delta \Omega /(2I) $. Together the contributions of the
 'resistive' (continuous vortex current) regions A and the 'capacitive'
 vortex trap (accumulation) regions B give Eq. (3).

The long term offset $\Delta \dot \Omega /\dot \Omega$ is observed
 to relax as a linear function of time:
\begin{equation}
\frac{\Delta \dot \Omega (t)}{\dot \Omega} =
\frac{I_{A}}{I}(1-\frac{t}{t_{g}})
\end{equation} 
The constants in this description of the observed long term 
$\Delta \dot \Omega (t)$ are labeled following the 
model for Vela (Alpar et al. 1984b). The time between glitches $t_{g}$ 
is the time it takes the spindown rate $\dot \Omega = N_{ext}/I$
 determined by the external torque to replenish the glitch induced offset 
$\delta \Omega$ in $\omega$:
\begin{equation}
t_{g}=\delta \Omega /\mid \dot \Omega \mid, 
\end{equation}
and
\begin{equation}
\frac{\Delta \Omega}{\Omega} = (\beta + \frac{1}{2})
(\frac{\Delta \dot \Omega}{\dot \Omega})\frac{\delta \Omega }{\Omega}
\end{equation}
where $\beta = {I_{B}}/{I_{A}}$.
Using Eqns (6)-(9), the long term second derivative of $\Omega$ to be 
observed between glitches is:
\begin{equation}
\ddot{\Omega} = \frac{I_{A}}{I} \frac{{\dot\Omega}^2}{\delta\Omega}
=(\beta + 1/2){(\Delta\dot\Omega/\dot\Omega)_{-3}}^2 / {(\Delta\Omega/\Omega)_{-6}} 
({\dot\Omega}^2 / \Omega ) .
\end{equation}
 This is equivalent to the positive 
"anomalous" braking index 
\begin{equation}
n = (\beta + 1/2){(\Delta\dot\Omega/\dot\Omega)_{-3}}^2 / 
{(\Delta\Omega/\Omega)_{- 6}}.
\end{equation}
The time to the next glitch can be expressed as 
\begin{equation}
t_{g}= 2 \times 10^{-3} {(\Delta\Omega/\Omega)_{- 6}}/[(\beta + 1/2)
{(\Delta\dot\Omega/\dot\Omega)_{-3}}] \tau_{sd}
\end{equation}
where $\tau_{sd} = \Omega/(2|\dot\Omega|)$ is the characteristic 
dipole spindown time.

We will show, in the next section, that the "anomalous" 
braking index behaviour of older pulsars is consistent 
with this model, indicating that all pulsars older than Vela experience 
glitches with $\Delta \Omega /\Omega > 10^{-7}$ 
and the universal interglitch behaviour described by Eqs(10) and (12). 
The hypothesis that all pulsars conform to this 
glitch behaviour model developed for the Vela pulsar was first applied 
to Geminga (Alpar, 
 {\"O}gelman $\&$ Shaham, 1993). Its universal application 
and implications for energy dissipation were introduced by Alpar 
(1998a,b, 2001). 

The significance of identifying this universal behaviour is that it implies 
a lower bound to the lag $\omega$ between crust and superfluid:
$\omega>\delta \Omega$
since the superfluid's loss of rotation rate at glitches should not overshoot
 the lag $\omega=\Omega_{s}-\Omega_{c}$.
 This lower bound in turn leads to a lower bound in the energy dissipation rate.

\section{Anomalous Braking Indices, Glitches and Interglitch Behaviour}

Braking indices were measured, at various degrees of accuracy as the 
data permitted, from 8 (excluding the Crab and Vela pulsars) out of 18 
glitching pulsars studied by Lyne, Shemar $\&$ Graham-Smith (2000),
 and from 9 (excluding the Vela pulsar) out of 11 glitching 
southern pulsars studied by Wang et al. (2000).
Some of these pulsars are common to both surveys.
 We exclude the Crab 
and Vela pulsars in the present work because detailed postglitch and 
interglitch data and fits exist for these pulsars; indeed the long term 
interglitch behaviour of the Vela pulsar provides the prototype
 dynamical behaviour that we are searching for in pulsars  older
 than the Vela pulsar. For three pulsars common to both
 surveys, PSR J 1341-6220, PSR J 1709-4428 and
 PSR J 1801-2304, Wang et al. (2000) quote $\ddot\Omega$
 measurements, while Lyne, Shemar $\&$ Graham-Smith (2000)
 quote upper limits to $\ddot\Omega$ for two of these pulsars.
 Thus there are now published $\ddot\Omega$ measurements
 for 14 out of 23 glitching pulsars excluding the Crab and Vela pulsars.   
 We have tabulated 10 of these according to the significance of error bars. 

In addition, Johnston $\&$ Galloway (1999) have obtained braking indices 
for 20 pulsars to demonstrate the method they proposed, applying 
Eq. (2) to rotation frequency and spindown rate measurements at 
two different epochs. These pulsars were not known glitching pulsars, 
and they were not observed to glitch during these observations. 
Anomalous braking indices were found for all 20 pulsars, with negative 
values in 6 pulsars
 and positive values in the rest.
Of the data in the Johnston and Galloway sample, we shall take into 
consideration 
 those data sets for which the quoted errors in the braking
 index are less than 
the quoted value, so that there is no ambiguity in the sign of the 
braking index. 
 With this criteria, we study 18 pulsars, 5 with negative and 13 
with positive braking indices. From two of these pulsars Johnston 
and Galloway reported two distinct data sets. Thus our sample contains 20 
determinations of the braking index from 18 pulsars.   
 Johnston $\&$ 
Galloway (1999) have interpreted the positive anomalous braking indices 
as due to interglitch recovery, without evoking a specific model. 
They interpreted the negative braking indices as reflecting
 an unresolved glitch during their observation time spans.
 All glitches result in long term decrease of the spindown
 rate, i.e. a negative step, an increase in the absolute
 value, of the rate of spindown. 
Since the pulsars were not monitored continuously, a glitch occurring 
between two timing observations would lead to a negative $\ddot\Omega$ 
inference, equivalent to a negative braking index. 

\section{Braking Indices of Pulsars Not Observed to Glitch}

We start our analysis with the braking indices measured by Johnston $\&$ 
Galloway (1999) from pulsars that were not observed to glitch, 
proceeding to the glitching pulsars in the next section. 
 All glitches bring about a sudden negative change $\Delta 
\dot{\Omega}$ in $\dot{\Omega}$, that is, a fractional
increase $\Delta\dot{\Omega}/\dot{\Omega}$ by 10$^{-3}$-10$^{-2}$ in the 
spindown rate. If the unresolved glitch happens in a
timespan of length t$_i$, the offset $\Delta \dot{\Omega}$ in the 
spindown rate will mimic a negative second derivative of
the rotation rate, $\ddot{\Omega} = \Delta \dot{\Omega}$/t$_i$.
Let us first elaborate on the statistical analysis of the
 negative braking index pulsars as those suffering an unobserved
 glitch during a gap within the timespan of the observations,
 following the analysis of Johnston $\&$ Galloway (1999) and using,
 as these authors did, the statistical glitch parameters of
 Alpar $\&$ Baykal (1994). The probability that
 pulsar i has one glitch during the timespan t$_i$ of the
 observations is given by the Poisson distribution 
\begin{equation}
 P (1 ; \lambda_i) = \lambda_i~ exp( - \lambda_i)
\end{equation}
where the parameter $ \lambda_i $ is given by
\begin{equation}
\lambda_{i} = \frac {t_{i}}{t_{g,i}}
\end{equation}
and $t_{g,i}$ is the time between glitches for pulsar i.
 To derive $t_{g,i}$ with Eq.(8), one needs to know the
 decrease  $\delta \Omega_i$ in superfluid rotation rate
 at the previous glitch. In this sample of pulsars from
 which glitches have not been observed we estimate the
 value of $\delta \Omega_i$ by making two alternative hypotheses 
about the constancy of average glitch parameters among
 pulsars older than the Vela pulsar and equating the parameters
 to their average values for the Vela pulsar glitches.
 Under the first hypothesis $\delta \Omega$ is assumed
 to be constant for all pulsar glitches, and
 is set equal to $< \delta \Omega > _{Vela}$, 
the average value inferred for the Vela pulsar glitches:
\begin{eqnarray}
{\delta \Omega_i}^{(1)} & = & < \delta \Omega > _{Vela}\\
{\lambda_i}^{(1)}  & = 
& \frac {t_i {|\dot{\Omega}|}_i }{ < \delta \Omega > _{Vela}}.
\end{eqnarray}
Under the second hypothesis, $\delta \Omega / \Omega $ is assumed 
to be constant for all glitches of pulsars older than the Vela pulsar.
 Johnston $\&$ Galloway (1999) adopted this hypothesis,
 taking the value estimated by Alpar $\&$ Baykal (1994)
 from glitch statistics, which agrees with the range of
 values of $\delta \Omega / \Omega $ inferred for the Vela pulsar glitches, 
\begin{eqnarray}
<\delta \Omega / \Omega>_{i}^{(2)} & = & 1.74 \times 10^{-4}     \\
\lambda_i^{(2)}  & = & 5.75 \times 10^{3} \frac {t_i |\dot{\Omega}|_i }
{ \Omega_i } = 2.87 \times 10^{-3} \frac {t_i}{\tau_{i, 6}}. 
\end{eqnarray}
Here t$_i$ is in years and $\tau_{i, 6}$ is the dipole
 spindown age of pulsar i in units of 10$^6$ years.
 Table 1 gives the values of ${\lambda_i}^{(1)}$ 
and ${\lambda_i}^{(2)}$. The corresponding probabilities P (1 ; $\lambda_i$) for 
an (unobserved) glitch to fall within the observation timespan
 devoted to pulsar i, or, equivalently, pulsar i mimicking a negative
 second derivative, are quite low for either hypothesis,
 while the probabilities P ( 0 ; $\lambda_i) \cong 1 $
 for no glitch occurring within the observation timespan of pulsar i,
 or, equivalently, a positive anomalous braking index being measured for pulsar i. 
The probability that 5 out of the 18 pulsars' 20 data sets 
 sampled have had unresolved glitches within the observation timespans,
 so that they have negative anomalous second derivatives, is given by 
\begin{equation}
 P (5 ; \lambda^{(j)}) = (\lambda^{(j)})^{5} ~ exp( - \lambda^{(j)})/5!
\end{equation}
where 
\begin{equation}
\lambda^{(j)} = \sum_{i=1}^{20} {\lambda_i}^{(j)}  
\end{equation}
for the hypotheses j = 1, 2. 
The index in this runs over all data sets, since 2 of the 8 pulsars have 
two independent data sets each in the sample of 
Johnston $\&$ Galloway (1999). 
 We find that 

\begin{eqnarray}
\lambda^{(1)} & =1.33 & \\
P (5 ; \lambda^{(1)}) & =0.0092 & \\
\lambda^{(2)} & =3.11 & \\
P (5 ; \lambda^{(2)}) & =0.11 & .
\end{eqnarray}

This means that hypothesis (2) is likely to be true,
 since it gives a total expected number of glitches
 falling within observation timespans to be 3.11
 against the number 5 implied by this interpretation of
 negative braking indices,
 as Johnston $\&$ Galloway (1999) noted.
 With hypothesis (1) the expected number of glitches is
 $\lambda^{(1)}$ = 1.33 
 and 5 glitches within observation timespans has a
 lower 
$P (5 ; \lambda^{(1)})  =0.0092 $ 
probability so this hypothesis is not favored.
The same conclusion was reached by Alpar $\&$ Baykal (1994) 
on the basis of statistics of large pulsar glitches: 
with the hypothesis (1), that $\delta \Omega $ is roughly 
constant in all pulsars older than Vela, 
the statistics implied $<\delta \Omega > = 0.0188$, 
which does not  agree with 
$<\delta \Omega >_{Vela} = 0.0094$

 In Table 1 the fractional changes in the spindown rate in
 the five unobserved glitches are given, 
as inferred from the negative braking indices by
 Johnston $\&$ Galloway (1999) according to:
\begin{equation}
(\frac {\Delta\dot{\Omega}_i}{\dot{\Omega}_i})_{missed} = 
\frac {\ddot{\Omega}_i t_i}{\dot{\Omega}_i} 
=\frac{n_{i} \dot \nu _{i} t_{i}}{\nu _{i}}.
\end{equation} 
These values
 $\Delta \dot \Omega _{i} / \dot \Omega _{i} \sim 
10^{-4}-10^{-3}$, are typical for glitching pulsars, 
 all measured values of $\Delta \dot \Omega / \dot \Omega $ 
for the Crab and Vela pulsars' large or small glitches are in 
the $10^{-4} - 10^{-3}$ range. Using these estimated 
values, and Equation (9) we can also estimate  
$\Delta \Omega / \Omega $ for the missed glitches. 
We assume that 
$\beta $ has similar values, 
$\beta \sim 0(1)$, in all glitching pulsars. 
Thus, taking $\beta+1/2 =1$, 
\begin{equation}
(\frac{\Delta \Omega}{\Omega})_{missed} 
=(\frac{\Delta \dot \Omega}{\dot \Omega})_{missed}
<\frac{\delta \Omega}{\Omega}>=
1.74 \times 10^{-4} (\frac{\Delta \dot \Omega}{\dot \Omega})_{missed}.
\end{equation}
We tabulate in Table 1 the estimated sizes of the missed glitches, 
$(\Delta \Omega / \Omega )_{max} \sim  (0.2-7) 10^{-7}$
 for the 5 pulsars with negative
 braking indices.
 Finally, we can check if glitches of the estimated magnitudes  
would have been missed in Johnston and Galloway's observations. 
The minimum glitch magnitude that can be detected through 
a mismatch of timing fits before and after the glitch is:
\begin{equation}
(\frac{\Delta \Omega _{i}}{\Omega _{i}})_{detectable} 
=\frac{\dot \nu _{i} t_{i}}{\nu _{i}}, 
\end{equation} 
which is of the order of $10^{-6}-10^{-5}$ for the data sets on these 5 pulsars. 
Thus, the interpretation that these negative braking indices indeed reflect
 undetected glitches is consistent with standard glitch models. 

The pulsars having positive braking indices reported by Johnston $\&$ Galloway 
(1999) must have been observed during interglitch relaxation. 
None of these pulsars have experienced a glitch during the observation 
time spans $t_{i}$. The values of $\lambda _{i}^{(1)}$ and 
$\lambda _{i}^{(2)}$ in \\
 Table 1 show that the probabilities 
P ( 0 ; $\lambda^{(j)}_{i}$)
 for no glitch occurring within the observation timespan 
of pulsar i are close to 1 under either hypothesis.
The positive interglitch $ \ddot \Omega $ values of 
these pulsars are related to the parameters
 of the previous glitch through Eq(10). 
Using this equation, we obtain the range of $\beta $ values corresponding 
to the range of positive braking indices, 
n = 2.5-50 quoted by Johnston and Galloway. Thus we expect $\beta $= 2.5 - 50, if 
$\Delta \Omega / \Omega = 10^{-6}$, 
$\Delta \dot \Omega / \dot \Omega = 10^{-3}$, 
while $\beta = 0.25 - 5 $ is obtained if 
$\Delta \Omega / \Omega = 10^{-7}$ and 
$\Delta \dot \Omega / \dot \Omega = 10^{-3}$. 
 
\section{Pulsars with Anomalous Braking Indices and Observed Glitches}

In this section, we discuss the pulsars which have been observed to glitch, 
and for which observations of anomalous braking indices, which are 
not noise artefacts, exist. So far, samples of such 
pulsars have been reported by Lyne, Shemar $\&$ Graham-Smith (2000) 
and by Wang et al. (2000). 

Many of these pulsars have exhibited multiple glitches, 
of varying magnitudes, from 
$\Delta \Omega / \Omega \sim 10^{-9}$ to
$\Delta \Omega / \Omega \sim 10^{-6}$. 
Reported $\ddot \Omega $ measurements are both negative and 
positive.
 Quoted errors in $\ddot \Omega $ are typically 
very large; especially among the negative $\ddot \Omega $ 
values reported.
 There is only one instance of 
a negative $\ddot \Omega $ with low error, 
among the glitching pulsars
reported by Wang et al. (2000),
$\ddot \Omega = -1.2318 \pm 0.019 \times 10^{-25}$ rad Hz s$^{-1}$   
in one particular epoch of observations for PSR J 1614-5047. 
  The epoch of this measurement does not
 coincide with  the only data set containing a glitch from this 
pulsar. We select from the data reported by Wang et al. (2000) and 
by Lyne, Shemar $\&$ Graham-Smith (2000) all those glitches 
with $\Delta \Omega / \Omega \ge 10^{-7}$. Among the 10 large glitches, 
with $\ddot \Omega $ measurements at or immediately following the glitch, 
8 glitches have positive $\ddot \Omega $ measurements. The two large glitches 
with subsequent negative second derivative measurements are 
from PSR J 1105-6107, with 
$\ddot \Omega = -3.078 \pm 0.314 \times 10^{-26}$ rad Hz s$^{-2}$, and 
PSR J 1801-2451, with 
$\ddot \Omega = -8.796 \pm 3.769 \times 10^{-26}$ rad Hz s$^{-2}$. 
As has been observed from the Vela pulsar, in postglitch relaxation 
after a large glitch, smaller glitches, with 
$\Delta \Omega / \Omega \sim 10^{-9}$ can sometimes occur. 
There is a possibility that the postglitch data set following 
these two glitches contains unresolved small glitches, 
$\Delta \Omega / \Omega \le 10^{-9}$, which determines the 
second derivative, and makes comparison with the model 
impossible. We therefore include only the 8 large 
glitches with measured positive post-glitch 
frequency second derivatives.      
 Observed values of 
$\Delta \Omega / \Omega $, 
$\Delta \dot \Omega / \dot \Omega $ and 
$\ddot \Omega $ are given in Table 2. We evaluate these quantities in terms  
of the "standard" interglitch response model given in Eqs (6)-(12). 
The extracted values of $\beta $, $\delta \Omega $, 
$\delta \Omega / \Omega $ and $t_{g}$ are also given in Table 2.     
The values of $\beta $ derived here are comparable to $\beta $ values inferred 
from model fits 
to the interglitch relaxation of the Vela pulsar with an 
exception for PSR 1709-4428. The 
$\delta \Omega $ values vary between 0.057 $\times 10^{-2}$ and 1.48 
$\times 10^{-2}$ , while 
$\delta \Omega / \Omega $  variation is less limited; 
the $\delta \Omega / \Omega $ values are similar to the values
 inferred for the sample of 
negative braking index pulsars (Table 1), and also to 
$<\delta \Omega / \Omega > \cong 1.74 \times 10^{-4} $
 inferred earlier from statistics. 

\section{Discussion}

We find that in glitching pulsars with measured 
braking indices, in the current sample, all pulsars 
exhibit positive second derivatives corresponding to interglitch 
recovery with model parameters similar to those obtained in 
detailed fits to interglitch behaviour of the Vela pulsar 
with the vortex creep model. This extends similar conclusions 
already reported on the basis of earlier, limited data. 

The main uncertainty in comparing these glitching pulsars 
with the model lies in the interpretation of the observed jumps 
$\Delta \dot \Omega / \dot \Omega $ in spindown rate. These 
glitch observations do not resolve the glitch occurrence time or 
the time dependence of $\Delta \dot \Omega $. 
Thus the quoted 
$\Delta \dot \Omega / \dot \Omega $ values may contain contributions 
from transients. The second derivatives characteristic of interglitch recovery
 are linked to only the long term offset in 
$\Delta \dot \Omega / \dot \Omega $, after the transients are over.
The transients and long-term contributions to 
$\Delta \dot \Omega / \dot \Omega $ are comparable in the 
Vela pulsar. Thus, this uncertainty introduces errors in 
$\beta $ estimates by factors of order 1. 

We have also explored Johnston and Galloway's measurements of positive and 
negative anomalous braking indices from a sample of pulsars which were not
 observed to glitch. These authors suggested that negative braking 
indices are due to the negative $\Delta \dot \Omega $ signs of 
unresolved glitches, while positive braking indices correspond to interglitch
 recovery. We have applied these suggestions specifically in the context of 
the phenomenology of Vela pulsar glitches and interglitch recovery.
The glitch model parameters are once again in agreement with parameters 
obtained for the Vela pulsar. 

Thus, on the basis of data from all pulsars with measured 
reliable secular anomalous braking indices, including both glitching pulsars
 and those without observed glitches, we conclude that 
pulsars older than the Vela pulsar experience glitches 
which are similar to the Vela pulsar's glitches. 
The interval between glitches is 
\begin{equation}
t_{g}=\frac{\delta \Omega}{\Omega}\frac{\Omega}{\mid \dot \Omega \mid}
\cong 2<\frac{\delta \Omega}{\Omega}> \tau _{sd}
\cong 3.5 \times 10^{-4} \tau _{sd}.
\end{equation} 
The last equality is on the basis of the strong indication, 
both from analysis of the statistics of all large glitches 
(Alpar $\&$ Baykal, 1994) and also from the analysis in this paper of the 
specific samples of pulsars with anomalous braking indices. 

A particularly interesting implication of the universality 
of glitch behaviour is the provision of a lower limit to 
the rate of energy dissipation due to vortex creep in neutron stars. 
As developed first by Alpar et al., (1984b), this energy dissipation rate 
is
\begin{equation}
\dot E _{diss}= I_{p}~ \omega \mid \dot \Omega \mid
\end{equation}
where $I_{p} \cong 10^{43}$ gm-cm$^{2}$ is the moment of inertia of 
the pinned inner crust superfluid where vortex creep takes place, and 
$\omega $ is the lag in the rotation rates between this inner crust
 superfluid and the observed outer crust. This expression is actually quite 
model independent. Upper limits on $\dot E_{diss}$ are obtained from 
observations of thermal X-ray emission from 
PSR B 1929+10 (Slowikowska et al. 2005, Alpar et al. 1987)) 
and PSR B 0950+58 
(Zavlin et al. 2004, Becker et al. 2004).
 The glitch related decrease in the rotation rate of the
 superfluid, $\delta \Omega $, provides a lower limit in the energy dissipation rate, 
since $\delta \Omega < \omega $: 
\begin{equation}
\dot E _{diss} = I_{p} \omega \mid \dot \Omega \mid 
    > I_{p} \delta \Omega \mid \dot \Omega \mid 
    \cong \frac{I_{p}}{I}<\frac{\delta \Omega}{\Omega}>I\Omega \dot \Omega
    \cong 1.7 \times 10^{-6} \dot E _{rot},
\end{equation}        
taking the moment of inertia ratio $I_{p}/I = 10^{-2}$ 
and $<\delta \Omega / \Omega >~ = 1.74 \times 10 ^{-4}$. 

Neutron stars older than a few $10^{6}$ yrs
 will have cooled to luminosities below \\ 
$\sim 10^{31}$ erg s$^{-1}$. The neutron star is then kept re-heated by 
energy dissipation. Thus $\dot E_{diss}$
 is actually a lower limit to the thermal 
luminosity of an old neutron star.
 The corresponding lower limit to the surface blackbody
 temperature of the neutron star is: 
\begin{equation}
T_{s}\ge 2.2 \times 10^{-4} \dot E_{rot} ^{1/4} R_{6} ^{-1/2} 
\end{equation}

For a radio pulsar spinning down as a pure dipole, 
extrapolating with the parameters of the Vela pulsar, 
\begin{equation}
\dot E_{rot} = 8.6 \times 10^{32} I_{45} t_{6} ^{-2}
\end{equation}
where $t_{6}$ is the age in $10^{6}$ years. Thus the lower 
limit becomes:
\begin{eqnarray}
L_{th} \cong \dot E_{diss} \ge 1.5 \times 10^{27} I_{45} t^{-2}_{6} \\
T_{s} \ge 3.8 \times 10^{4} I_{45}^{1/4} R_{6}^{-1/2} t_{6}^{-1/2}
\end{eqnarray} 

Unfortunately this limit on the black-body temperature
 is in the UV band  
at an age of $10^{7}$ yrs. If the actual energy dissipation rate 
is close to the upper limits applied by the PSRs B 1929+10 and B 0950+58, 
we have  

\begin{eqnarray}
L_{th} \le 1.2\times 10^{30} (I_{p} \omega)_{43}  t_{6}^{-3/2} \\
T_{s}  \le
 2.0\times 10^{5}  (I_{p} \omega)_{43}^{1/4} R_{6}^{-1/2} t_{6}^{-3/8}
\end{eqnarray}

If a neutron star is spinning down under a more constant torque 
like for instance a propeller torque from a fallback disk, 
and if such spindown extends beyond the few $10^{6}$ years of the 
initial cooling era, the luminosity and surface 
temperature sustained by energy dissipation, according 
Eqs (32) and (33), might be observable.

{\bf{Acknowledgments}}

 We thank the anonymous referee 
for his/her comments that led to clarification of
several points.
 MAA thanks the Turkish Academy of Sciences and we thank the 
 Sabanc{\i} University 
Astrophysics and Space Forum for research support.

{\bf{References}}

\hspace{-1.1cm} Alpar M.A., Anderson P.W., Pines D., Shaham J.,
                1984a, ApJ., 276, 325

\hspace{-1.1cm} Alpar M.A., Anderson P.W., Pines D., Shaham J.,
                1984b, ApJ., 278, 791

\hspace{-1.1cm} Alpar M.A., Langer S.A., Sauls, J.A., 1985, ApJ.,
                 282, 533

\hspace{-1.1cm} Alpar M.A., Nandkumar R., Pines D.,
                1986, ApJ., 311, 197

\hspace{-1.1cm} Alpar, M.A., Brinkman, W., {\"O}gelman, H., 
                K{\i}z{\i}loglu, {\"U}., Pines, D., 
                1987, A$\&$A, 177, 101 

\hspace{-1.1cm} Alpar M.A., Cheng K.S., Pines D., 1989, ApJ.,
                346, 823

\hspace{-1.1cm} Alpar, M.A., Chau, H.F., Cheng, K.S.,
                and Pines, D. 1993, ApJ., 409, 345

\hspace{-1.1cm} Alpar, M.A., Baykal, A., 1994,
                 MNRAS, 269, 849

\hspace{-1.1cm} Alpar M.A., 1998a, in R.Buccheri,
               J. van Paradijs, and M.A. Alpar (eds.),
               The Many Faces of Neutron Stars, Proc. NATO-ASI,
                Kluwer, p59

\hspace{-1.1cm} Alpar, M.A., 1998b,
                Proc. COSPAR General Assembly, Birmingham 1996,
                Adv. Space Res. Vol.21, 159 (1998b) 

\hspace{-1.1cm} Alpar, M.A., {\"O}gelman, H., Shaham, J., 
                1993, A$\&$A, 273, L35

\hspace{-1.1cm} Alpar, M.A., 2001, Proc. NATO ASI, Elounda, Greece,
                eds. C. Kouveliotou, J.Ventura $\&$ J. van Paradijs, Kluwer  

\hspace{-1.1cm} Alpar, M.A., 2005, Proc. NATO ASI, Marmaris, Turkey, eds. 
                A.Baykal, S.K.Yerli, S.C.Inam, $\&$ S. Grebenev  

\hspace{-1.1cm} Arzoumanian Z., Nice D.J., Taylor J.H.,
                1994, 422, 671

\hspace{-1.1cm} Baykal A., Alpar M.A., Boynton P.E.,
                Deeter J.E., 1999, MNRAS, 306, 207

\hspace{-1.1cm} Becker, W., Weisskopf, M.C., Tennant, A.F.,
                Jessner, A., Dyks, J., 
                Harding, A.K. and Shuang, N.Z., 2004, ApJ 615, 908

\hspace{-1.1cm} Boynton P.E., Groth E.J., Hutchingson D.P.,
                Nanos G.P., Partridge R.B., Wilkinson D.T.,
                1972, ApJ.,  175, 217

\hspace{-1.1cm} Camilo, F., Kaspi, V.M., Lyne, A.G., Manchester, R.N.,
                 Bell, J.F., 
                 D'Amico, N., McKay, N.P.F. and Crawford, F.,
                 2000, ApJ 541, 367

\hspace{-1.1cm} Cordes J. M., Helfand D.J., ApJ.,
               1980, 239, 640

\hspace{-1.1cm} Cordes J.M., ApJ., 1980, 237, 216

\hspace{-1.1cm} Cordes J. M., Downs G. S.,
                ApJ. Sup., 1985, 59, 343

\hspace{-1.1cm} Cheng K.S., 1987a, ApJ., 321, 799

\hspace{-1.1cm} Cheng K.S., 1987b, ApJ., 321, 803

\hspace{-1.1cm} Cheng K.S., 1989, in {\"O}gelman H.,
                 van den Heuvel E.P.J., eds. NATO ASI
                 Ser. Vol. 262. Timing Neutron Stars,
                  Kluwerm Dordrecht. p.
                  503

\hspace{-1.1cm} D'Alessandro F., McCulloch P.M., Hamilton P.A.,
                Deshpande A.A., 1995, MNRAS,  277, 1033

\hspace{-1.1cm} Deshpande A.A., D'Alessandro F., McCulloch P.M.,
                1996, J.Astrophys. Astr.  17, 7

\hspace{-1.1cm} Downs G.S., Reichley P.E., 1983,
                Astrophys. J. Suppl. Ser.,
                 53, 169

\hspace{-1.1cm} Gouiffes C., Finley J.P.,
                {\"O}gelman H.,1992, ApJ, 394, 581

\hspace{-1.1cm} Groth E.J., 1975, Astrophys. J. Suppl. Ser.,
                 29, 443

\hspace{-1.1cm} Jones P.B., 1990, MNRAS,  246, 364

\hspace{-1.1cm} Johnston S., Galloway D., 1999,
                MNRAS, 306, L50

\hspace{-1.1cm} Kaspi V.M., Manchester R., Siegman B.,
               Johnston S., Lyne A.G., 1994,
               ApJ.,422, L83

\hspace{-1.1cm} Livingstone, M.A., Kaspi, V.M.,
                Gotthelf, E.V. and Kuiper, L., 2006, 
                astro-ph 0601530, submitted to ApJ

\hspace{-1.1cm} Lyne A.G., Pritchard R.S.,
               Smith F.G., ApJ., 1982, 260, 520

\hspace{-1.1cm} Lyne A.G., Pritchard R.S., Smith F.G.,
               1988, MNRAS, 223, 667

\hspace{-1.1cm} Lyne A.G., Pritchard R.S., Smith F.G.,
               1993, MNRAS, 265, 1003

\hspace{-1.1cm} Lyne A.G., Pritchard R.S., Smith F.G.,
              1996, Nature, 381, 497

\hspace{-1.1cm} Lyne A. G., Shemar S. L., Smith F. Graham.,
                2000, MNRAS, 315, 534

\hspace{-1.1cm} Manchester R.N., Peterson B.A., 1989,
               ApJ, 342, L23

\hspace{-1.1cm} Nagase F., Deeter J., Lewis W.,
               Dotani T., Makino F.,
               Mitsuda K., 1990, ApJ, 351, L13

\hspace{-1.1cm} Shemar S. L., Lyne A. G., 1996,
                MNRAS, 282, 677

\hspace{-1.1cm} Siegman B.C., Manchester R.N., Durdin J.M.,
                1993, MNRAS, 262, 449

\hspace{-1.1cm} Slowikowska, A., Kuiper, L. Hermsen, W., 2005, A$\&$A 434, 1097

\hspace{-1.1cm} Wang N., Manchester R. N., Pace R. T.,
               Bailes M., Kaspi V. M., Stappers
               B. W., Lyne A. G., 2000, MNRAS, 317, 843

\hspace{-1.1cm} Zavlin, V.E., Pavlov, G.G., 2004, ApJ 616, 452

 \newpage
\flushright
\begin{table}
\caption{Pulsars with positive or negative braking indices  $^{a} $}
\begin{center}
\begin{tabular}{| c| c| c| c| c| c| c| c| c| c| }

PSR B  & t(days) & $\Omega $ & $\dot \Omega_{17}$  &
$ \lambda^{1}  $ & $\lambda^{2}$ & 
$\frac{\Delta \dot \Omega }{  \dot \Omega } $ & $\frac {\Delta \Omega }{ \Omega }_{m} $ & 
$ \frac{\Delta \dot\Omega }{\dot \Omega } _{m} $ &
 n  \\
 & & rad Hz & rad Hz~$s^{-1}$ & $\times 10^{-2} $&$\times 10^{-2} $  
&$\times 10^{-3} $
&$\times 10^{-7} $ &$=\frac{\dot \Omega t_{i}}{ \Omega } $ & \\
 & & & & & & & 
&$\times 10^{-5} $ & \\

0114+58& 2271.1& 61.9& -35.3 & 7.37 & 6.43 & 0.11 &0.19 &1.1 &   -9.6$\pm$1.5 \\
0136+57& 4492.0&  23.1& -9.1 & 3.74 & 8.75 & 1.2 &2.1 &1.5 &  -81 $\pm$4.7 \\
0154+61& 4336.5&  2.7& -2.1  & 0.86 & 17.11 &             & & &   28 $\pm$14 \\
0540+23& 5543.5&  25.5& -16.0 & 8.16 & 17.28 &             & & &  11.1$\pm$ 8.6 \\
       & 5990.5&     &        & 8.82 & 18.67 &             & & & 11.81$\pm$ 0.12 \\
0611+22& 5541.5& 18.8& -33.4 & 16.98 & 48.84 &             & & & 20.1 $\pm$ 1.1 \\
0656+14& 2163.3& 16.3& -23.3 & 4.63 & 15.34 &             & & & 14.7 $\pm$ 1.4 \\
0740-28& 4245.2&  37.7& -38.0 & 14.83 & 21.25 &             & & & 17.7 $\pm$ 1.4 \\
       & 5827.2&     &        & 20.35 & 29.17 &             & & & 25.6 $\pm$ 0.8 \\
0919+06& 4521.7&  14.6&  -4.6 & 1.93 & 7.16 &             & & & 28.9 $\pm$ 4.1 \\
1221-63& 6661.3& 29.0& -6.6 & 4.06  & 7.56         & &    & &18.7 $\pm$12.3 \\
1719-37& 4824.0& 26.6& -12.2 &  5.42 & 11.02 & 3.5 &6.1 &1.9 &-183  $\pm$ 10 \\
1742-30& 1581.0& 17.1&  -5.0 & 0.72 & 2.26 & 0.52 &0.91 &0.39 &-132  $\pm$  5 \\
1829-08& 1541.0& 9.7& -9.5 & 1.34 & 7.50  &             & & & 2.5  $\pm$ 0.9 \\
1907+10& 5842.5& 22.1&  -2.1 & 1.10 & 2.70  &    & &         & 24   $\pm$ 17 \\
1915+13& 6080.5& 32.3& -11.9 & 6.67 & 11.16  &             & & & 36.08$\pm$ 0.48 \\
2000+32& 1381.0& 9.0& -13.6 & 1.72 & 10.30  & 4.1 &7.1 &1.8 &-226  $\pm$ 4.5 \\
2002+31& 6076.5& 3.0&  -1.0 & 0.58 & 10.64  &             & & & 23.3 $\pm$ 1.0 \\
2148+52& 2307.2& 18.9&  -5.7 & 1.21 & 3.45  &             & & & 49.6 $\pm$ 3.5 \\
2334+61& 2347.1& 12.7& -48.8 & 10.52 & 44.78  &             & & &  8.6 $\pm$ 0.13  \\
\end{tabular}
\end{center}
$^{a}$ Johnston $\&$ Galloway (1999)
\end{table}
\begin{table}
\begin{flushleft}
\caption{Observed parameters of glitching pulsars  }
\begin{center}
\begin{tabular}{| c| c| c| c| c| c| 
c| c| c| c|   }
PSR & $\Omega $  & $ \mid \dot \Omega _{-11} \mid$  & $\ddot \Omega _{-22}$   &
 $(\frac{\Delta \Omega}{\Omega })_{-6} $ & $(\frac{\Delta \dot \Omega}{\dot \Omega})_{-3} $ &
$\beta $ & t$_{g}$ & $\delta \Omega _{-2}$  &
$(\frac{\delta \Omega}{ \Omega}) _{-4}$ 
 \\
  &rad Hz & rad Hz s$^{-1}$ &rad Hz s$^{-2}$ & & & &days & &  \\
1048-583 $^{a}$ & 50.8 & 3.95 & 9.22 & 2.995 & 3.7 & 6.07    &1834  & 0.626  & 1.23 \\
         &      &      & 24.5 & 0.771 & 4.62 & 2.37    &865  & 0.296  & 0.582  \\
1341-6230 $^{a}$& 32.5 & 4.25 & 11.93& 0.99  & 0.7  & 42.87   &288  & 0.106  & 0.326 \\
         &      &      & 16.96& 1.636 & 3.3  & 4.08    &958  & 0.352  & 1.082 \\
1614-5047 $^{a}$& 27.1 & 5.8  & 21.99& 6.456 & 9.7  & 0.72    &2950  & 1.48   & 5.576 \\
1709-4428 $^{a}$& 61.32& 5.57 & 10.87& 2.012 & 0.2  & 1080.16 &119 & 0.057  & 0.09    \\
1730-3350 $^{b}$& 45.2 &2.76  &  6.28& 3.0   & 12.  & 0.27    &6154.1& 1.467& 3.24      \\
1740-3015 $^{b}$& 10.68&0.79  & 6.59 & 0.4   & 3.   & 4.51    &416.4 & 0.028& 0.26          \\
         &      &      & 3.95 & 0.6   & 2.   & 9.63    &936.9 & 0.063& 0.58           \\     
1803-2137 $^{b}$& 47.12&4.77  & 8.04 & 4.0   & 9.2  & 0.28    &6373.1& 2.62 & 5.56         \\
1801-2451 $^{a}$& 50.31& 5.15 & 25.07& 1.998 & 4.85 & 3.54    &1153    & 0.51   & 1.02    \\
1803-2137 $^{a}$& 47.  & 4.7  & 18.00& 3.2   &10.7  & 1.16    &2894  & 1.17   & 2.5     \\  
\end{tabular}
\end{center}
$^{a}$ Wang et al. 2000 \\
$^{b}$ Lyne et al. 2000 \\ 
\end{flushleft}
\end{table}


\end{document}